\documentclass[epj,nopacs]{svjour}
\usepackage{graphicx}
\usepackage{amsmath,amssymb}
\usepackage{array}
\usepackage{arydshln}
\usepackage{color}
\usepackage{cite}

\allowdisplaybreaks[2]

\newcommand{\be}{\begin{equation}}
\newcommand{\ee}{\end{equation}}
\newcommand{\bea}{\begin{eqnarray}}
\newcommand{\eea}{\end{eqnarray}}
\newcommand{\beq}{\begin{equation}}
\newcommand{\eeq}{\end{equation}}
\newcommand{\beqa}{\begin{eqnarray}}
\newcommand{\eeqa}{\end{eqnarray}}
\newcommand{\ba}{\begin{array}}
\newcommand{\ea}{\end{array}}

\newcommand{\nn}{\nonumber}

\newcommand{\cG}{{\cal G}}
\newcommand{\cO}{{\cal O}}

\newcommand{\ord}[1]{\mathcal{O}({#1})}

\newcommand{\lsim}{\stackrel{<}{_\sim}}
\newcommand{\gsim}{\stackrel{>}{_\sim}}
\definecolor{nicered}{rgb}{0.7,0.1,0.1}
\newcommand{\cB}{{\cal B}}

\newcommand{\no}{\nonumber}
\newcommand{\yuk}{{\lambda}}

\newcommand{\Qbar}{\overline{Q}}
\newcommand{\Dbar}{\overline{D}}
\newcommand{\Ebar}{\overline{E}}
\newcommand{\Lbar}{\overline{L}}

\newcommand{\Mbar}{\overline{M}}

\newcommand{\identity}{1 \hspace{-.085cm}{\rm l}}
\newcommand{\sw}{s^2_{\rm W}}

\renewcommand{\makeheadbox}{}
\vspace{-5cm}
\title{Minimal Flavour Violation and Beyond}
\author{Gino Isidori\inst{1,2} \and David M. Straub\inst{3}}
\institute{INFN, Laboratori Nazionali di Frascati, Via E.~Fermi 40, 00044 Frascati, Italy 
\and CERN, Theory Division, 1211 Geneva 23, Switzerland
\and Scuola Normale Superiore and INFN, Piazza dei Cavalieri 7, 56126 Pisa, Italy}

\date{}
\abstract{ 
We review the formulation of the Minimal Flavour Violation (MFV) hypothesis in the quark sector, as well 
as some ``variations on a theme'' based on smaller flavour symmetry groups and/or less minimal breaking terms. 
We also review how these
hypotheses can be tested  in $B$ decays and by means of other flavour-physics observables. 
The phenomenological consequences of MFV are discussed both in general terms, 
employing a general effective theory approach, and in the specific context of the Minimal Supersymmetric 
extension of the SM. 
\PACS{
      {PACS-key}{describing text of that key}   \and
      {PACS-key}{describing text of that key}
     } 
} 

\begin{document}

\maketitle

\section{Introduction}

The Standard Model (SM) is an effective theory valid up to some still undetermined cutoff scale $\Lambda$. The gauge hierarchy problem suggests that this scale should be in the TeV region currently being probed at LHC, where some new physics (NP) should appear. If this NP interacts directly or indirectly with the SM particles, it necessarily contribute to flavour-violating processes. Since there is no exact flavour symmetry in the SM (the flavour symmetry of the gauge sector is broken by the Yukawa interactions), one cannot assume an exact flavour symmetry in the NP model: some breaking would unavoidably appear at the quantum level.

On the other hand, the excellent agreement of flavour data with the SM predictions implies that, for generic flavour violation, the scale $\Lambda$ cannot be low. For example, $K^0$-$\bar K^0$ mixing alone sets a bound on $\Lambda$ of roughly $10^5$~TeV for generic $\Delta S=2$ flavour-violating effective operators of dimension six~\cite{Isidori:2010kg}.
Consequently, insisting on a solution to the hierarchy problem by TeV-scale NP, we are forced to conclude that the flavour structure of the NP is highly non-generic, especially in the quark sector. On general grounds, a NP flavour structure
able to satisfy the existing tight constraints requires two main ingredients: i) a large flavour symmetry, and ii) small symmetry-breaking terms. Given the flavour-violating structure present in the SM, the most restrictive assumption to {\em protect} in a consistent way quark-flavour mixing beyond the SM is to assume that the flavour symmetry is the one present in the SM in the limit of vanishing Yukawa couplings (namely a $U(3)^3$ quark-flavour symmetry~\cite{Chivukula:1987py}) and that the two 
quark Yukawa couplings are the only two irreducible symmetry-breaking terms.
As a result of this hypothesis of {\em Minimal Flavour Violation} (MFV), which can naturally be formulated with the language of effective theories \cite{D'Ambrosio:2002ex}, non-standard contributions in flavour-changing neutral current (FCNC) transitions turn out to be suppressed to a level consistent with experiments even for $\Lambda \sim$~few TeV. 

Similarly to the violation of flavour, the violation of the CP symmetry is a challenge for any NP theory. The non-observation of electric dipole moments (EDMs) of fundamental fermions puts stringent lower bounds on the scale $\Lambda$, assuming generic (flavour-conserving) CP-violating NP interactions. This problem persists even in theories with MFV since the minimal breaking of flavour does not preclude the presence of {\em flavour blind} CP-violating phases. 

It should also be kept in mind that, while the MFV principle would naturally explain the absence so far of any signals of flavour violation beyond the SM, it is not a theory of flavour, in the sense that there is no explanation for the observed hierarchical structure of the Yukawa couplings. Finally, it is worth to stress that while the MFV hypothesis is the most efficient way to suppress NP contribution to flavour-violating processes, it is by no means the only symmetry + symmetry-breaking pattern allowed by present data.

This review is organized as follows. In section~\ref{sec:mfv} we discuss the MFV principle in its effective field theory formulation, analyzing in detail the different implementations in theories with one or two Higgs doublets.
In section~\ref{sec:mfv2} we discuss some possible variations around the MFV hypothesis, addressing in particular the 
issue of flavour-blind phases, emphasizing the special role of the third-generation Yukawa couplings (non-linear formulations), and addressing the compatibility with Grand Unified Theories. 
In section~\ref{sec:MSSM} we analyze the implementation of the MFV hypothesis in the specific context of the Minimal 
Supersymmetric extension of the SM (MSSM). In section~\ref{sec:EffSUSY} we discuss possible variations of the MFV hypothesis relevant in the 
so-called 
{\em split-family} SUSY framework: a MSSM with a large mass gap 
between the first two generations of squarks and the third one. 
We review in particular the split-family  framework with a  $U(2)^3$ flavour symmetry, 
which provides an interesting attempt to explain (at least in part) the observed Yukawa hierarchies,
addressing at the same the absence of large deviations from the SM in flavour- and CP-violating 
 observables.

\section{Minimal Flavour Violation}
\label{sec:mfv}

The largest set of unitary, global field rotations in the quark sector commuting with the SM gauge symmetry is~\cite{Chivukula:1987py}
\begin{equation}
\cG_q = U(3)_{Q_L} \times U(3)_{U_R} \times U(3)_{D_R} \,.
\label{eq:Gq}
\end{equation}
As mentioned, we cannot promote this global flavour symmetry to be an exact symmetry beyond the SM, since it is already broken within the SM by the Yukawa interactions,
\be
\label{eq:SMY}
- {\cal L}^{\rm SM}_{\rm Yukawa}=Y_d^{ij} {\bar Q}^i_{L} \phi D^j_{R}
 +Y_u^{ij} {\bar Q}^i_{L} \tilde\phi U^j_{R}
+ {\rm h.c.}  
\ee
(where $\tilde\phi=i\tau_2\phi^\dagger$).
The MFV hypothesis consists in assuming
that $Y^d$ and $Y^u$ are the only sources of flavour symmetry breaking also in the NP model.  To implement and interpret this hypothesis in a consistent way, we can assume that $\cG_q$ is a good symmetry and promote $Y^{u,d}$ to be non-dynamical fields (spurions) with non-trivial transformation properties under $\cG_q$:
\begin{equation}
Y^u \sim (3, \bar 3, 1)\,,\qquad
Y^d \sim (3, 1, \bar 3)\,.
\end{equation}
If the breaking of the symmetry occurs at very high energy scales, at low-energies we would only be sensitive to the background values of the $Y$, i.e. to the ordinary SM Yukawa couplings.  Employing the effective theory language, an effective theory satisfies the MFV criterion in the quark sector if all higher-dimensional operators, constructed from SM and $Y$ fields, are formally invariant under the flavour group $\cG_q$~\cite{D'Ambrosio:2002ex}.\footnote{The notion of MFV can also be extended to the lepton sector; however, in this case there is not a unique way to define the minimal sources of flavour symmetry breaking if one wants to accommodate non-vanishing neutrino masses. We postpone a discussion about this point to section~\ref{sec:GUT}.}

As mentioned in the introduction, not only flavour-violating interactions but also flavour-conserving CP-vio\-lating interactions
provides a serious challenge to NP models at the TeV scale. Following the approach of Ref.~\cite{D'Ambrosio:2002ex}, 
this problem can be circumvented enlarging the symmetry to $\cG_q \times $CP and assuming that the Yukawa couplings are the only symmetry breaking terms of both $\cG_q$ and CP. 

According to the MFV criterion one should in principle consider operators with arbitrary powers of the (dimensionless) Yukawa fields. However, a strong simplification arises by the observation that all the eigenvalues of the Yukawa matrices are small, but for the top one (and possibly the bottom one, see later), and that the off-diagonal elements of the CKM matrix are very suppressed. Working in the basis
\beq\label{speint}
Y^d=\lambda_d,\ \ \ Y^u=V^\dagger\lambda_u,
\eeq
where $\lambda_{d,u}$ are diagonal,
\beq\label{deflamd}
\lambda_d={\rm diag}(y_d,y_s,y_b),\ \ \
\lambda_u={\rm diag}(y_u,y_c,y_t),
\eeq
and neglecting the ratio of light quark masses over the top mass, we have
\be
\left[  Y^u (Y^u)^\dagger \right]^n_{i\not = j} ~\approx~
y_t^{2n} V^*_{ti} V_{tj}~.
\label{eq:basicspurion}
\ee
As a consequence, including high powers of the the Yukawa matrices amounts only to a redefinition of the overall factor in (\ref{eq:basicspurion}) and the the leading $\Delta F=2$ and $\Delta F=1$ FCNC amplitudes get exactly the same CKM suppression as in the SM:
\begin{eqnarray}
&&  \mathcal{A}(d^i \to d^j)_{\rm MFV} =  (V^*_{ti} V_{tj}) \mathcal{A}^{(\Delta F=1)}_{\rm SM}
\left[ 1 + a_1 \frac{ 16 \pi^2 M^2_W }{ \Lambda^2 } \right], \no \\  
&& \mathcal{A}(M_{ij}-{\Mbar_{ij}})_{\rm MFV}  =  (V^*_{ti} V_{tj})^2
 \mathcal{A}^{(\Delta F=2)}_{\rm SM} \times \no \\
&& \qquad\qquad\qquad\qquad  \times \left[ 1 + a_2 \frac{ 16 \pi^2 M^2_W }{ \Lambda^2 } \right]~,
\label{eq:FC}
\end{eqnarray}
where the $\mathcal{A}^{(i)}_{\rm SM}$ are the SM loop amplitudes and the $a_i$ are $\mathcal{O}(1)$ parameters.
The  $a_i$  depend on the specific operator considered but are flavour independent. This implies the same relative correction in $s\to d$, $b\to d$, and  $b\to s$ transitions of the same type. In the minimal set-up, where CP is also a good symmetry of the theory in absence of Yukawa couplings, 
the  $a_i$ are real parameters.\footnote{As pointed out in \cite{Mercolli:2009ns}, 
this statement is not fully correct since the $a_i$ could have non-vanishing
flavour-blind phases proportional the Jarlskog invariant $J_{\rm CP}={\rm det}[ Y^u (Y^u)^\dagger,  Y^d (Y^d)^\dagger]$.
However, the smallness of $J_{\rm CP}$ implies that these phases play a negligible role in flavour-violating observables 
(unless enhanced by unnaturally large coefficients).}

\subsection{Universal UT and MFV bounds on the effective operators}

\begin{figure*}[t]
\begin{center}
\vskip -3.8 true cm
\includegraphics[width=85mm]{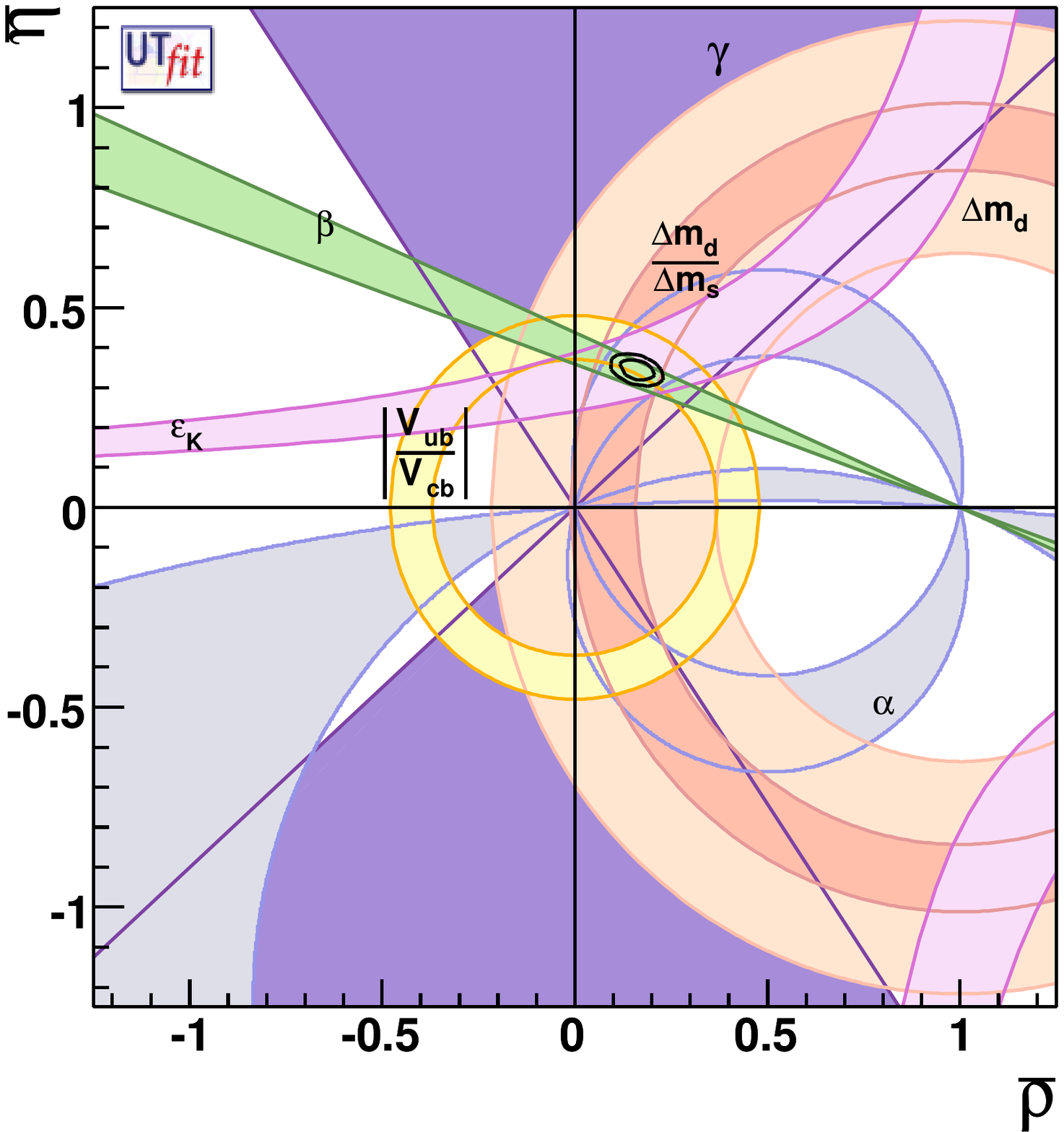}
\includegraphics[width=85mm]{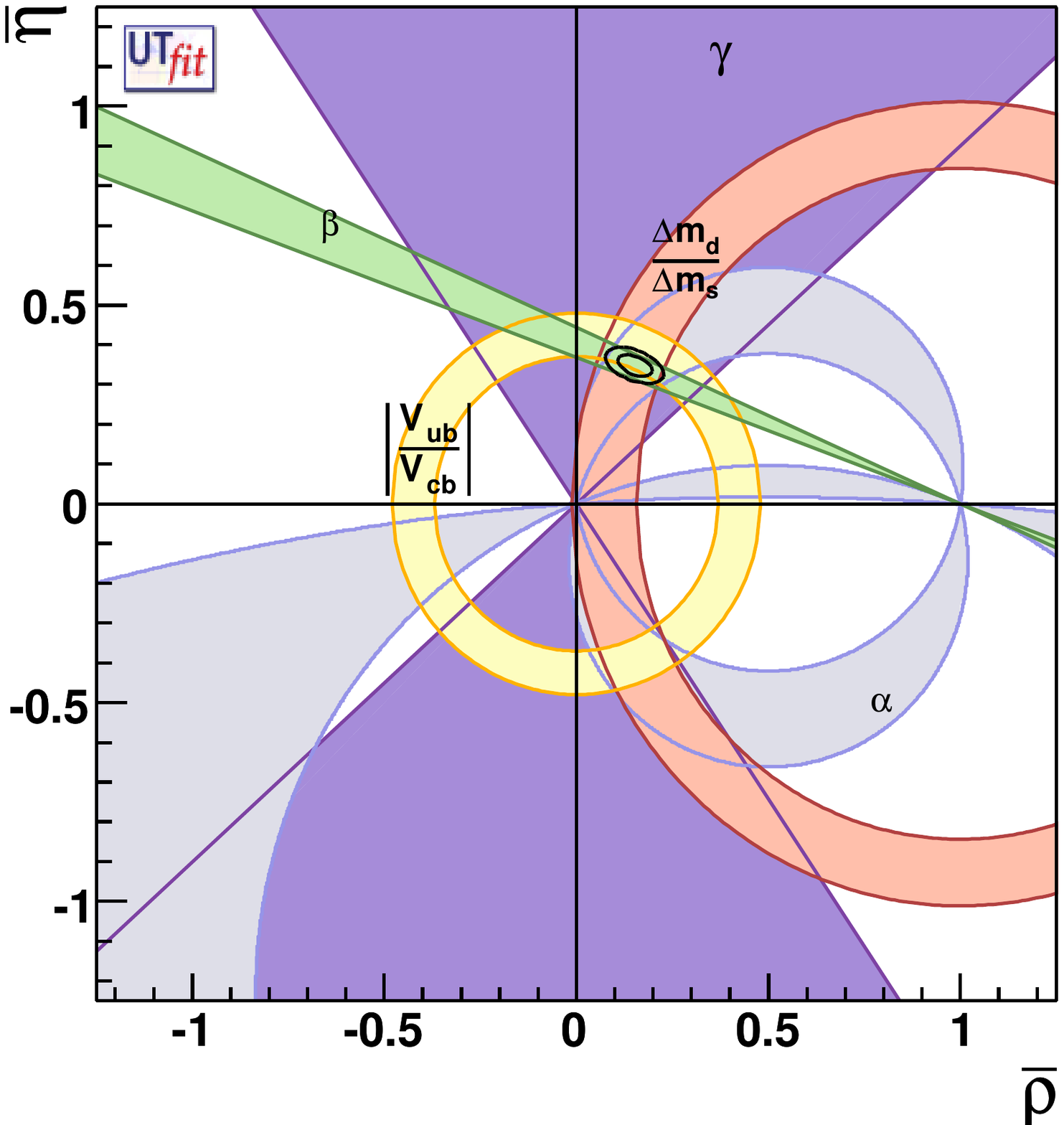}\\
\vskip 0.5 true cm
\caption{\label{fig:UTfits} Fit of the CKM unitarity triangle within the SM (left) and 
in generic extensions of the SM satisfying the MFV hypothesis (right), from Ref.~\cite{Bona:2006sa}. }
\end{center}
\end{figure*}

As originally pointed out in Ref.~\cite{Buras:2000dm}, 
within the MFV framework several of the constraints used to determine the 
CKM matrix (and in particularly the unitarity triangle) are not affected by NP. 
In this framework, NP effects are negligible not only in tree-level processes 
but also in a few clean observables sensitive to loop effects, such as 
the time-dependent CPV asymmetry in $B_d \to J/\Psi K_{L,S}$. Indeed 
the structure of the basic flavour-changing 
coupling in Eq.~(\ref{eq:FC}) implies that the weak CPV phase of 
$B_d$--$\bar B_d$ mixing is arg[$(V_{td}V_{tb}^*)^2$],
exactly as in the SM. 
The determination of the unitarity triangle using only these clean observables 
(denoted Universal Unitarity Triangle) is shown in  
Fig.~\ref{fig:UTfits}.\footnote{~The Unitairty Triangle shown on the left 
plot of Fig.~\ref{fig:UTfits} includes also the $\Delta M_{B_s}/\Delta M_{B_d}$ 
constraint, assuming this ratio is not modified with respect 
to the SM. This condition holds only in the so-called 
constrained MFV scenario of~Ref.~\cite{Buras:2000dm} (see Sect.\ref{sec:CMFV}).}
This construction provides a
natural (a posteriori) justification of why no NP effects have 
been observed in the quark sector: by construction, most of the clean 
observables measured at $B$ factories are insensitive to NP effects 
in this framework.

\begin{table*}[t]
\begin{center}
\begin{tabular}{l|c|l}
Operator & ~Bound on $\Lambda$~  & ~Observables \\
\hline\hline
$H^\dagger \left( \Dbar_R Y^{d\dagger}  Y^u Y^{u\dagger}
  \sigma_{\mu\nu} Q_L \right) (e F_{\mu\nu})$ 
& ~$6.1$~TeV & ~$B\to X_s \gamma$, $B\to X_s \ell^+ \ell^-$\\
$\frac{1}{2} (\Qbar_L  Y^u Y^{u\dagger} \gamma_{\mu} Q_L)^2 
\phantom{X^{X^X}_{iii}}$
& ~$5.9$~TeV & ~$\epsilon_K$, $\Delta m_{B_d}$, $\Delta m_{B_s}$ \\   
$H_D^\dagger \left( \Dbar_R  Y^{d\dagger} 
Y^u Y^{u\dagger}  \sigma_{\mu\nu}  T^a  Q_L \right) (g_s G^a_{\mu\nu})$
&~$3.4$~TeV & ~$B\to X_s \gamma$, $B\to X_s \ell^+ \ell^-$\\
$\left( \Qbar_L Y^u Y^{u\dagger}  \gamma_\mu
Q_L \right) (\Ebar_R \gamma_\mu E_R)$  
& ~$2.7$~TeV & ~$B\to X_s \ell^+ \ell^-$, $B_s\to\mu^+\mu^-$ \\
$~i \left( \Qbar_L Y^u Y^{u\dagger}  \gamma_\mu
Q_L \right) H_U^\dagger D_\mu H_U$
&~$2.3$~TeV 
&~$B\to X_s \ell^+ \ell^-$, $B_s\to\mu^+\mu^-$\\
$\left( \Qbar_L Y^u Y^{u\dagger}  \gamma_\mu Q_L \right) 
( \Lbar_L \gamma_\mu L_L)$
&~$1.7$~TeV & ~$B\to X_s \ell^+ \ell^-$, $B_s\to\mu^+\mu^-$\\
$\left( \Qbar_L Y^u Y^{u\dagger}  \gamma_\mu Q_L
\right) (e D_\mu F_{\mu\nu})$
&~$1.5$~TeV & ~$B\to X_s \ell^+ \ell^-$\\
\end{tabular}
\end{center}
\caption{\label{tab:MFV} Bounds on the scale of new physics (at 95\%
  C.L.) for some representative $\Delta F=1$~\cite{Hurth:2008jc} and  
$\Delta F=2$~\cite{Bona:2007vi} MFV operators 
(assuming effective coupling $\pm 1/\Lambda^2$), and corresponding 
observables used to set the bounds.}
\end{table*}

In  Table~\ref{tab:MFV} we report a few representative 
examples of the bounds on the higher-dimen\-sio\-nal operators
in the MFV framework.
As can be noted, the built-in CKM suppression
leads to bounds on the effective scale of new physics 
not far from the TeV. These bounds are very similar to the 
bounds on flavour-conserving operators derived by precision electroweak tests. 
This observation reinforces the conclusion that a deeper study of 
rare decays is definitely needed in order to clarify 
the flavour problem: the experimental precision on the clean 
FCNC observables required to obtain bounds more stringent 
than those derived from precision electroweak tests
(and possibly discover new physics) is typically
in the $1\%-10\%$ range.

Although MFV seems to be a natural solution to the flavour problem, it should be stressed that  we are still far from having proved the validity of this hypothesis from data (in the effective theory language we can say that there is still room for sizable new sources of flavour symmetry breaking beside the SM Yukawa couplings~\cite{Feldmann:2006jk}).  A proof of the MFV hypothesis can be achieved only with a positive evidence of physics beyond the SM exhibiting the flavour-universality pattern (same relative correction in $s\to d$, $b\to d$, and $b\to s$ transitions of the same type) predicted by the MFV assumption.  While this goal is quite difficult to be achieved, the MFV framework is quite predictive and could easily be falsified. For instance, no significant enhancement over the SM predictions 
is expected in direct CP asymmetries in singly-Cabibbo suppressed charm decays: as in the SM case, the asymmetries are suppressed by 
${\rm Im}(V_{ub}^*V_{cb}/V_{ud}^*V_{us})$~\cite{Isidori:2011qw}. As a result, if the recent evidence of direct-CP violation
in the charm system~\cite{Aaij:2011in} will be 
established to be incompatible with the SM expectation, this would rule out not only the SM, but would also be a clear signal 
against MFV models.

\subsection{Comparison with other approaches}
\label{sec:CMFV}

The idea that the CKM matrix rules the strength of FCNC 
transitions also beyond the SM has become a very popular 
concept in the recent literature and has been implemented 
and discussed in several works 
(see e.g.~Refs.~\cite{Ali:1999we,Buras:2000dm}).

It is worth stressing that the CKM matrix 
represents only one part of the problem: a key role in
determining the structure of FCNCs  is also played  by quark masses, 
or by the Yukawa eigenvalues. In this respect, the MFV 
criterion illustrated above provides the maximal protection 
of FCNCs (or the minimal violation of flavour symmetry), 
since the full structure of Yukawa matrices is preserved. 
At the same time, this criterion is based on a renormalization-group-invariant 
symmetry argument. Therefore, it can be implemented 
independently of any specific hypothesis about the dynamics 
of the new-physics framework. The only two assumptions are:
i) the flavour symmetry and its breaking sources; 
ii) the number of light degrees of freedom of the theory 
(identified with the SM fields in the minimal case).
 
This model-independent structure does not hold in 
most of the alternative definitions of MFV models 
which can be found in the literature. For instance, 
the definition of Ref.~\cite{Buras:2003jf} 
(denoted constrained MFV, or CMFV)
contains the additional requirement that only the 
effective FCNC operators which play a significant 
role within the SM are the only relevant ones 
also beyond the SM. 
This condition is realized within weakly coupled 
theories at the TeV scale with only one light Higgs doublet, such as the MSSM 
with small $\tan\beta$ and small $\mu$ term.
However, it does not hold in other frameworks, such as 
technicolour/composite models (see e.g.~\cite{Kagan:2009bn,Redi:2011zi})
 or the MSSM with large 
$\tan\beta$ and/or large $\mu$ term,
whose low-energy phenomenology can still be described 
using the general MFV criterion discussed 
in Sect.~\ref{sec:mfv}.

\subsection{\boldmath MFV at large $\tan\beta$}
\label{sec:largetanb}

The flavour group in Eq.~(\ref{eq:Gq}) can be decomposed as 
\bea
\cG_q &=& U(1)^3 \times SU(3)_{Q_L} \times SU(3)_{U_R} \times SU(3)_{D_R} \no \\
&=& U(1)^3 \times SU(3)^3_{q}~.
\label{eq:Gq2}
\eea
If the Yukawa Lagrangian contains more than one single Higgs field it is natural to treat separately the 
breaking of the $U(1)$ groups and that of $SU(3)^3_{q}$.  To this purpose, we note that 
two of the $U(1)$ charges can can be identified with the baryon number 
and the (quark) hypercharge, which are not broken by the Yukawa interaction
even in the SM case.  The third Abelian group, $U(1)_{\rm PQ}$, can be identified as 
a group under which only the three $D^i_R$ fields are charged (with the same charge), 
while $U^i_R$ and $Q^i_L$ are neutral (Peccei-Quinn symmetry). Only this Abelian symmetry is explicitly 
broken by the Yukawa interaction in the one-Higgs doublet case. 

In the two-Higgs doublet case, assigning different $U(1)_{\rm PQ}$ charges to the two 
Higgs fields ($\phi_U$ neutral and $\phi_D$ with opposite charge to $D_R$), 
we can write a $U(1)_{\rm PQ}$-invariant Yukawa interaction:
\begin{equation}\label{eq:LY2}
- \mathcal{L}^{\rm 2HDM}_{\rm Yukawa}  = Y^d_{ij}~{\Qbar_{Li}}\phi_D
D_{Rj}
+Y^u_{ij}~{\Qbar_{Li}} \phi_U U_{Rj}
+{\rm h.c.}
\end{equation}
This interaction prevents tree-level FCNCs and implies that $Y^{u,d}$ are the only sources of $SU(3)^3_{q}$ breaking appearing in the Yukawa interaction (similar to the one-Higgs-doublet scenario). Consistently with the MFV hypothesis, we can then assume that $Y^{u,d}$ are the only relevant sources of $SU(3)_{q}^3$ breaking appearing in all the low-energy effective operators.  This is sufficient to ensure that flavour-mixing is still governed by the CKM matrix, and naturally guarantees a good agreement with present data in the $\Delta F =2$ sector. However, the extra symmetry of the Yukawa interaction allows us to change the overall normalization of $Y^{u,d}$ with interesting phenomenological consequences in specific rare modes. These effects are related only to the large value of bottom Yukawa, and indeed can be found also in other NP frameworks where there is no extended Higgs sector, but the  bottom Yukawa coupling is of order one~\cite{Kagan:2009bn}.

Assuming the Lagrangian in Eq.~(\ref{eq:LY2}), the normalization of the Yukawa couplings is controlled by the ratio of the vacuum expectation values of the two Higgs fields, or by the parameter
\be
\tan\beta = \langle \phi_U\rangle/\langle \phi_D\rangle~.
\ee
For $\tan\beta\gg1 $ the smallness of the $b$ quark and $\tau$ lepton masses can be attributed to the smallness of $1/\tan\beta$ rather than to the corresponding Yukawa couplings.  As a result, for $\tan\beta\gg1$ we cannot anymore neglect the down-type Yukawa coupling. In this scenario the determination of the effective
low-energy Hamiltonian relevant to FCNC processes
involves the following three steps:
\begin{itemize}
\item{} construction of the gauge-invariant basis of
dimension-six operators (suppressed by $\Lambda^{-2}$)
in terms of SM fields and two Higgs doublets;
\item{} breaking of ${\rm SU}(2)\times {\rm U}(1)_Y$ and
integration of the  $\cO(M_H^2)$ heavy Higgs fields;
\item{} integration of the $\cO(M_W^2)$ SM degrees
of freedom (top quark and electroweak gauge bosons).
\end{itemize}
These steps are well separated if we assume the
scale hierarchy $\Lambda \gg M_H \gg M_W$.
On the other hand, if $\Lambda \sim M_H$, the first
two steps can be joined, resembling the
one-Higgs-doublet scenario discussed before.
The only difference is that now, at large $\tan\beta$,
$\yuk_D$ is not negligible and this leads to enlarge
the basis of effective dimension-six operators.
From the phenomenological 
point of view, this implies the breaking of the strong MFV 
link between $K$- and $B$-physics FCNC amplitudes 
occurring in the  one-Higgs-doublet case \cite{D'Ambrosio:2002ex}. 

\medskip 

A more substantial modification of the one-Higgs-dou\-blet
case occurs if we allow sizable sources of 
 ${\rm U}(1)_{\rm PQ}$ breaking. It should be pointed out 
that the ${\rm U}(1)_{\rm PQ}$ symmetry cannot be exact:
it has to be broken at least in the scalar potential
in order to avoid the presence of a massless pseudoscalar Higgs.
 Even if the breaking of ${\rm U}(1)_{\rm PQ}$ and $SU(3)^3_{q}$ are decoupled, the presence of ${\rm U}(1)_{\rm PQ}$ breaking sources can have important implications on the structure of the Yukawa interaction, especially if $\tan\beta$ is large~\cite{Hall:1993gn,Blazek:1995nv, Isidori:2001fv,D'Ambrosio:2002ex}.  We can indeed consider new dimension-four operators such as
\begin{equation}
 \epsilon~ \Qbar_L  Y^d D_R  \tilde \phi_U
\qquad {\rm or} \qquad
 \epsilon~ \Qbar_L  Y^uY^{u\dagger} Y^d D_R  \tilde \phi_U~,
\label{eq:O_PCU}
\end{equation}
where $\epsilon$ denotes a generic MFV-invariant ${\rm U}(1)_{\rm
  PQ}$-breaking source. 
The effective Yukawa Lagrangian then assumes the form
\begin{eqnarray}
- \mathcal{L}^{\rm 2HDM}_{\rm Y~eff}  
 &=& \bar Q_L X_{d1} D_R \phi_D + \bar Q_L X_{u1} U_R \tilde\phi_D \no\\
&& + \bar Q_L X_{d2} D_R \tilde\phi + \bar Q_L X_{u2} U_R \phi_U +{\rm h.c.}~, \qquad
\label{eq:generalcouplings}
\end{eqnarray}
where 
\bea
X_{d1} &=& P_{d1}(Y_u Y_u^\dagger, Y_d Y_d^\dagger) \times Y_d~,  \\
X_{d2} &=& P_{d2}(Y_u Y_u^\dagger, Y_d Y_d^\dagger) \times Y_d~, \\
X_{u1} &=& P_{u1}(Y_u Y_u^\dagger, Y_d Y_d^\dagger) \times Y_u~, \\
X_{u2} &=& P_{21}(Y_u Y_u^\dagger, Y_d Y_d^\dagger) \times Y_u~, 
\eea
and $P_{i}(Y_u Y_u^\dagger, Y_d Y_d^\dagger)$ are generic polynomials 
of the two basic left-handed spurions
$Y_u Y_u^\dagger$ and $Y_d Y_d^\dagger$.
Since we are free to re-define the two basic spurions $Y_u$ and $Y_d$, without loss of generality 
we can define them to be the flavour structures appearing in $X_{d1}$ and $X_{u2}$. Then expanding 
the remaining non-trivial polynomials in powers of $Y_u^\dagger Y_u$ and $Y_d^\dagger Y_d$ 
leads to~\cite{D'Ambrosio:2002ex,Buras:2010mh}
\bea
X_{d1} &=& Y_d~,  \nn \\
X_{d2} &=& \epsilon_{0} Y_d + \epsilon_{1} Y_d  Y_d^\dagger Y_d                    
+  \epsilon_{2}  Y_u Y_u^\dagger Y_d + \ldots~,  \nn \\
X_{u1} &=& \epsilon^\prime_{0} Y_u + \epsilon^\prime_{1}  Y_u Y_u^\dagger Y_u 
+  \epsilon^\prime_{2}  Y_d Y_d^\dagger Y_u + \ldots~, \nn \\
X_{u2} &=& Y_u~.
\label{eq:XMFVgen} 
\eea

  Even if $\epsilon \ll 1 $, the product
$\epsilon \times \tan\beta$ can be $\mathcal{O}(1)$, inducing large
corrections to the down-type Yukawa sector:
\begin{align}
 \epsilon~ \Qbar_L  Y^d D_R  \tilde \phi_U
 ~\stackrel{\text{vev}}{\longrightarrow}~
& \epsilon~  \Qbar_L  Y^d D_R   \langle \tilde \phi_U \rangle
\\
=\ & (\epsilon\times\tan\beta)~  \Qbar_L  Y^d D_R   \langle \phi_D \rangle~.
\end{align}
As discussed first in specific supersymmetric frameworks,
for $\epsilon\times \tan\beta = \cO(1)$ the ${\rm U}(1)_{\rm PQ}$-breaking
terms induce $\cO(1)$ corrections to the down-type Yukawa
couplings~\cite{Hall:1993gn}, the 
CKM matrix elements~\cite{Blazek:1995nv},
and the charged-Higgs 
couplings~\cite{Carena:1999py,Degrassi:2000qf,Carena:2000uj}.
Moreover, sizable FCNC couplings of the down-type quarks to the
heavy neutral Higgs fields 
are allowed~\cite{Hamzaoui:1998nu,Babu:1999hn,Isidori:2001fv}. 
All these effects can be taken into account to all orders
with a proper re-diagonalization of the effective Yukawa 
interaction in Eq.~(\ref{eq:generalcouplings}).

\begin{figure}[t]
\begin{center}
\includegraphics[width=.45\textwidth]{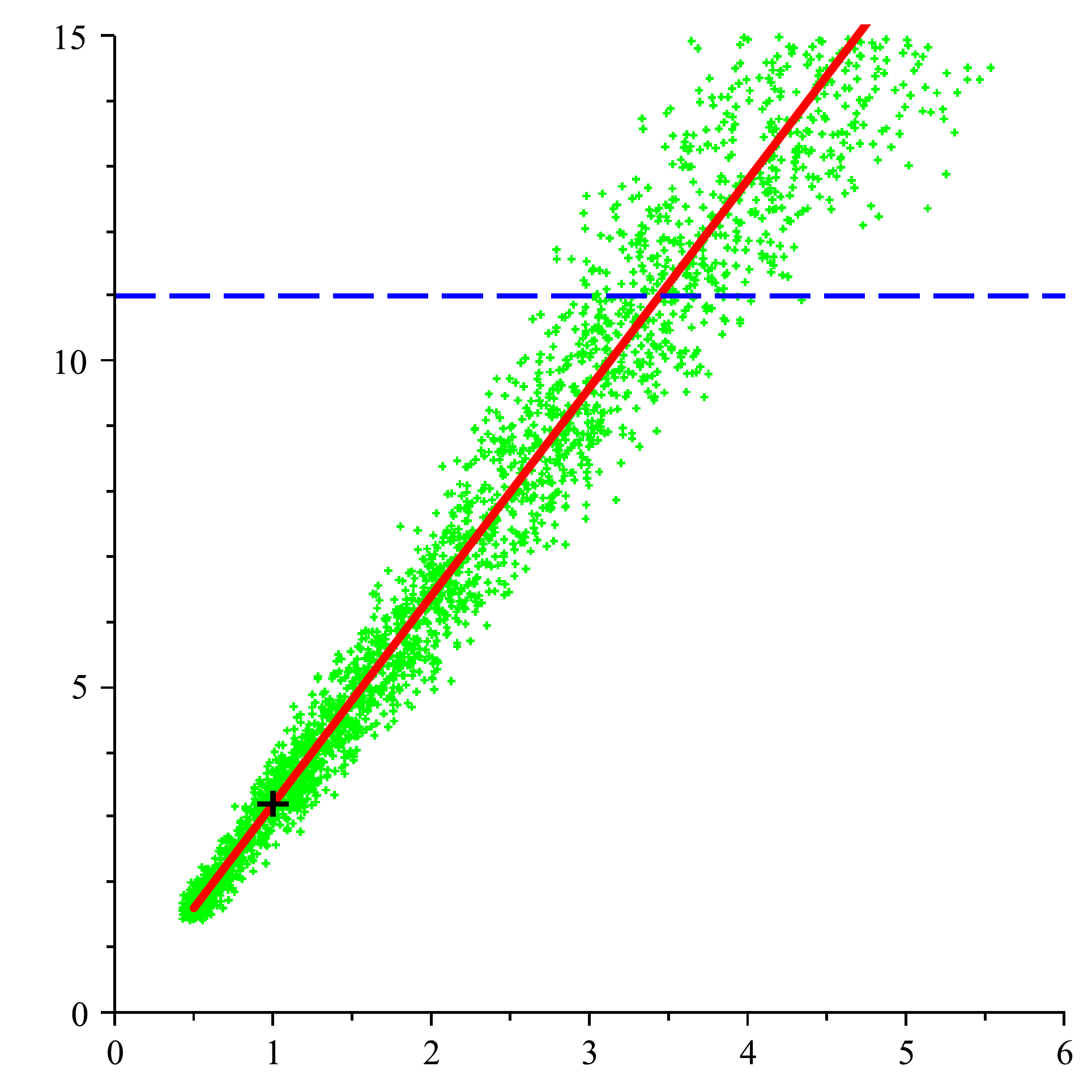}
\vskip -7.8 true cm 
\hskip -2.8 true cm  $10^9 \times \cB(B_{s}\to \mu^+\mu^-)$ \\
\vskip  6.1 true cm
\hskip  4.5 true cm  $10^{10} \times \cB(B_{d}\to \mu^+\mu^-)$ \\
\vskip  1.0 true cm
\caption{\label{fig:Bll}
Correlation between $\cB(B_s\to\mu^+\mu^-)$ and $\cB(B_d\to\mu^+\mu^-)$
in presence of Higgs-medited FCNC amplitudes respecting the MFV hypothesis.
The continuos red line indicates the central value of the correlation, while the 
green points take into acount the uncertainties in $|V_{ts}|$ and $|V_{td}|$. 
The black cross denotes the SM prediction. 
The horizontal dashed line indicates the combined $95\%$ CL upper limit from LHCb~\cite{Aaij:2011rj} and CMS~\cite{Chatrchyan:2011kr}.}
\end{center}
\end{figure}

Since the $b$-quark Yukawa coupling becomes $\mathcal{O}(1)$, the large-$\tan\beta$ regime is particularly interesting for helicity-suppressed observables in $B$ physics, such as the rare decays 
$B_{s,d}\to \ell^+\ell^-$, $B \to \ell \nu$, and $B\to X_s \gamma$.

The most striking signature could arise from $B_{s,d}\to \ell^+\ell^-$, whose decay rates could be substantially 
enhanced over the SM expectations even taking into account all the tight constraints from the other observables.  
The channels of this type where the experiments reach the best sensitivity compared to the SM expectations  
are $B_{s,d}\to\mu^+\mu^-$. 
Despite the recent improved limits from CDF~\cite{Aaltonen:2011fi}, LHCb~\cite{Aaij:2011rj} and CMS~\cite{Chatrchyan:2011kr},  
there is still substantial room for enhancements over the SM predictions~\cite{Buras:2010mh}
\bea
\cB(B_d\to\mu^+\mu^-)_{\rm SM} &=& (1.1\pm 0.15)\times 10^{-10}~,  \no \\
\cB(B_s\to\mu^+\mu^-)_{\rm SM} &=& (3.7\pm 0.4)\times 10^{-9}~.
\eea 
The correlation of $\cB(B_s\to\mu^+\mu^-)$ and $\cB(B_d\to\mu^+\mu^-)$ 
in a generic model with Higgs-medited FCNC amplitudes respecting the MFV hypothesis is shown in Fig.~\ref{fig:Bll}. 
An enhancement of both modes respecting the MFV relation $\Gamma(B_{s}\to \ell^+\ell^-)/\Gamma(B_{d}\to \ell^+\ell^-) \approx |V_{ts}/V_{td}|^2$  would be an unambiguous signature of MFV at large $\tan\beta$~\cite{Hurth:2008jc}.

As far as $B \to \ell \nu$  is concerned, in the limit of exact ${\rm U}(1)_{\rm PQ}$ symmetry the tree-level charged-Higgs amplitude
interfere destructively with the SM one~\cite{Hou:1992sy}: assuming  $m_H > m_W \times \tan\beta $, this implies a  suppression (typically in the $10-50\%$ range) of the decay rate with respect to its SM prediction.
However, for sizable ${\rm U}(1)_{\rm PQ}$-breaking terms  ( $|\epsilon \times \tan\beta| \gsim 1$), the sign of the 
correction is not unambiguosly determined and rate enhancements up to $50\%$ cannot be excluded~\cite{Blankenburg:2011ca}.
Potentially measurable effects in the $10-30\%$ range are expected also in $B\to X_s \gamma$~\cite{Carena:1999py}.

\section{Beyond the minimal set-up}
\label{sec:mfv2}

\subsection{MFV with flavour-blind phases}
\label{sec:FBP}

The breaking of the flavour group $\cG_q$ and the breaking of the discrete CP symmetry are not necessarily related, since generic MFV models can still contain flavour blind phases~\cite{Ellis:2007kb,Mercolli:2009ns,Kagan:2009bn}. Because of the experimental constraints on electric dipole moments (EDMs), which are generally sensitive to such flavour-diagonal phases~\cite{Mercolli:2009ns,Paradisi:2009ey}, in this more general case the bounds on the scale of new physics are substantially higher with respect to the ``minimal'' case, where the Yukawa couplings are assumed to be the only breaking sources of both symmetries~\cite{D'Ambrosio:2002ex}.

However, there are concrete examples of MFV models where flavour blind phases lead to observable effects in $B_d$ and $B_s$ mixing, or CP asymmetries in $B$ decays, without violating the EDM bounds. These include the MSSM with complex trilinear couplings of third generation squarks \cite{Altmannshofer:2008hc,Paradisi:2009ey} and a two Higgs doublet model with flavour blind phases in Yukawa interactions or the Higgs potential \cite{Buras:2010mh,Buras:2010zm}.

The correlation of CP-violating effects in the three observable $\Delta F=2$ systems ($B_{d,s}$ and $K$ meson mixing) 
is a powerful test of possible flavour-blind phases in a MFV framework.
At small $\tan\beta$ there is only one relevant $\Delta F=2$  operator:
\be
(\Qbar_L  Y^u Y^{u\dagger} \gamma_{\mu} Q_L)^2 
\ee
Since this operator is Hermitian, its coupling must be real and no deviations are expected in the 
$\Delta F=2$ sector compared to the case without flavour-blind phases. 
The situation changes if $\tan\beta$ is large. In this case, thanks to the large bottom Yukawa coupling, 
additional operators with the insertion of $Y_d$ break 
the universality between $K$ and $B$ systems. In the limit where we 
can neglect the strange-quark Yukawa coupling, 
the extra CPV induced by flavour blind phases is equal in $B_s$-$\bar B_s$ and $B_d$-$\bar B_d$ mixing and does not enter $K^0$-$\bar K^0$ mixing~\cite{Kagan:2009bn}.
This can be understood from an approximate $U(2)^3$ symmetry in the large $\tan\beta$ case (see section~\ref{sec:u23}).

Neglecting the strange-quark Yukawa coupling is usually a good approximation, but for the specific case of Higgs-mediated 
amplitudes with sizable  ${\rm U}(1)_{\rm PQ}$-breaking terms. In this case operators containing right-handed light quarks cannot 
be neglected and break the correlation between CPV in the $B_s$ and $B_d$ systems. In the limit where these 
scalar FCNCs are dominant, the modification of the $B_s$ mixing phase is $m_s/m_d$ larger with respect to the 
corresponding effect in the $B_d$ system~\cite{Buras:2010mh}.

While sizable corrections to  $B_s$ and $B_d$ mixing phases are possible in general, this is not the case in the specific 
realization of MFV with flavour-blind phases in the MSSM (even at large $\tan\beta$)~\cite{Altmannshofer:2009ne}. This can be traced back to the suppression in the MSSM of effective operators with several Yukawa insertions. Sizable couplings for these operators are necessary both to have an effective large CP-violating phase in $B_s$-$\bar B_s$ mixing and, at the same time, to evade bounds from other observables, such as $B_s\to \mu^+\mu^-$ and $B \to X_s \gamma$.

\subsection{\boldmath Non-linear representation, discrete subgroups, and gauging of $U(3)^3$}
\label{sec:NLMFV}

As stressed above, the MFV expansion relies on the smallness of the off-diagonal elements of the CKM matrix and the hierarchies 
between the Yukawa eigenvalues. It does not suffer from the fact of $y_t$ (and possibly $y_b$, at large $\tan\beta$) being sizable.
As explicitly shown in Eq.~(\ref{eq:basicspurion}), the effect of considering high powers in 
$y_t$ only modify the overall  strength of the basic flavour-violating spurion
\be
(V^\dagger \lambda_u^2 V)_{i\not = j}~.
\ee

An elegant implementation of the MFV hypothesis,
taking into account explicitly the special role the diagonal third-generation Yukawa couplings is obtained 
with a  non-linear realization of the flavour symmetry~\cite{Feldmann:2008ja,Kagan:2009bn}. 
Particularly interesting is the so-called GMFV case, where both  $y_t$  and $y_b$ are assumed to be of order one 
and their effects are re-summed to all orders~\cite{Kagan:2009bn}. As shown in~\cite{Kagan:2009bn}, the 
 flavour symmetry group surving after this resummation and linearly realised (with small breaking terms)
 is a $U(2)^3\times U(1)$ group. 
 
 Given the smallness of $y_{c,u}/y_t$ and 
$y_{s,d}/y_d$, as well as the smallness of the 
off-diagonal elements of the CKM matrix, the phenomenological predictions derived in the GMFV framework 
are not different from those obtained with the standard MFV expansion in $ Y^u$ and $Y^d$, 
provided the expansion is carried out up to the first non trivial terms. Indeed the difference between the 
GMFV predictions derived in~\cite{Kagan:2009bn} with respect to those obtained in~\cite{D'Ambrosio:2002ex},
employing the standard MFV expansion at large $\tan\beta$, can all be attributed  to the presence of 
flavour-blind phases in the GMFV set-up.  

A different scenario, recently analysed in Ref.~\cite{Alonso:2012jc}, is the case where the $SU(2)\times U(1)_Y$  electroweak symmetry 
is non-linealry realized, giving rise to a low-energy effective theory without a fundamental Higgs. In this case the 
situation is different with respect to the minimal case not because of the flavour symmetry, but because of the 
different power counting of the effective operators. This allow potentially interesting interference effects among 
terms that in the standard case are of different order in the derivative expansion. 

Going from the effective approach, where the Yukawas are treated as spurions, to a more fundamental level 
where the Yukawas are dynamical fields acquiring a non-trivial vacuum expectation value is a very difficult task.
First of all,  one has to face the  problem of constructing potentials 
giving rise to the peculiar hierarchial structure of the Yukawa:  attempts in this direction have recently 
been discussed in Ref.~\cite{Alonso:2011yg,Nardi:2011st}.  Moreover, a general problem that one encounters 
is the (unwanted) appearence of a large number of Goldstone bosons, associated to the spontaneous breaking 
of the large global continuos flavour symmetry~\cite{Albrecht:2010xh}. This problem could be avoided assuming that the fundamental
flavour symmetry is a suitable discrete subgroup of $\cG_q$, as discussed in~\cite{Zwicky:2009vt}, or trying to gauge the 
flavour symmetry, as discussed in~\cite{Grinstein:2010ve,Buras:2011wi}. However, as shown in these recent works, in both cases 
one finds a phenomenology quite different from the minimal MFV case illustrated before. Particularly interesting is the case 
of gauged flavour symmetries, where one expects a spectrum of massive gauge bosons with masses inversely proportional 
to the SM Yukawa couplings.

\subsection{MFV in Grand Unified Theories}
\label{sec:GUT}

The notion of MFV can also be extended to the lepton sector. However, in this case there is not a unique way to define the minimal sources of flavour symmetry breaking if one wants to accommodate non-vanishing neutrino masses. Indeed different versions of Minimal Lepton Flavour 
Violation (MLFV) have ben proposed in the literature, depending on how the irreducible breaking terms in the neutrino sector are 
identifed~\cite{Cirigliano:2005ck,Davidson:2006bd,Gavela:2009cd,Alonso:2011jd}. As for the quark sector, the key tool to possibly tests
these assumptions relies on the observation of possible correlations in the rate of neutral-current LFV processes, such as 
$\tau\to \mu\gamma$ and $\mu\to e\gamma$~\cite{Cirigliano:2005ck,Davidson:2006bd,Gavela:2009cd,Alonso:2011jd}. 

Once we accept the idea that flavour dynamics obeys a MFV
principle, it is interesting to ask if and how this is compatible with
Grand Unified Theories (GUTs), where quarks and leptons sit in the same
representations of a unified gauge group. This question has 
been addressed in Ref.~\cite{Grinstein:2006cg}, 
considering the exemplifying case of ${\rm SU}(5)_{\rm gauge}$.

Within ${\rm SU}(5)_{\rm gauge}$, the down-type singlet 
quarks ($D^{i}_{R}$) and the lepton doublets 
($L^i_{L}$) belong to the $\bar {\bf 5}$ representation; the quark doublet
($Q^i_{L}$), the up-type ($U^{i}_{R}$) and lepton singlets ($E^{i}_{R}$) 
belong to the ${\bf 10}$ representation, and finally 
the right-handed neutrinos ($\nu^i_{R}$) are singlet.
In this framework the largest 
group of flavour transformation commuting with 
the gauge group is
\be
{\mathcal G}_{\rm GUT} = 
{\rm U}(3)_{\bar 5} \times {\rm U}(3)_{10}\times {\rm U}(3)_1,
\ee 
which is smaller than the direct product 
of the quark and lepton flavour groups compatible 
with the SM gauge sector: ${\mathcal G}_q \times {\mathcal G}_l$,
where ${\mathcal G}_l =  {\rm U}(3)_{E_R} \times  {\rm U}(3)_{L_L}$.
We should therefore expect some violations 
of the MFV predictions, either in the quark sector, 
or in the lepton sector, or in both.

A phenomenologically acceptable description of the low-energy fermion 
mass matrices requires the introduction of at least four irreducible 
sources of ${\mathcal G}_{\rm GUT}$ breaking. From this point of view
the situation is apparently similar to the non-unified case: the four 
 ${\mathcal G}_{\rm GUT}$ spurions can be put in one-to-one 
correspondence with the low-energy spurions $Y^{u,d,e}$
plus the neutrino Yukawa coupling $Y^\nu$ 
(which is the only low-energy spurion in the neutrino sector 
assuming an approximately degenerate heavy $\nu_R$ spectrum). 
However, the smaller flavour group
does not allow the diagonalization of $Y^d$ and
$Y^e$ (which transform in the same way under ${\mathcal G}_{\rm GUT}$)
in the same basis. As a result, two additional mixing matrices 
can appear in the expressions for flavour changing rates \cite{Grinstein:2006cg}.
The hierarchical texture of the new mixing matrices is known
since they reduce to the identity matrix in the limit 
$(Y^e)^T = Y^d$. Taking into account this fact, and
analysing the structure of the allowed higher-dimensional operators, 
a number of reasonably  firm phenomenological consequences  
can be deduced~\cite{Grinstein:2006cg}: 
\begin{itemize}
\item 
There is a well defined limit in which the standard MFV 
scenario for the quark  sector is  fully recovered: 
$|Y_\nu|\ll 1$ and small $\tan \beta$. 
The upper bound on the neutrino Yukawa couplings implies 
an upper bound on the heavy neutrino masses ($M_\nu$). 
In the limit of a degenerate heavy neutrino spectrum, 
this bound is of about $10^{12}$ GeV.
For $M_\nu \sim  10^{12}$ GeV and small $\tan \beta$, 
deviations from the standard MFV pattern 
can be expected in rare $K$ decays but  not in $B$ physics.\footnote{~The 
conclusion that $K$ 
decays are the most sensitive probes of possible deviations from the  
strict MFV ansatz follows from the strong suppression of 
the $s \to d$ short-distance amplitude in the SM [$V_{td}V_{ts}^* =\cO(10^{-4})$],
and goes beyond the hypothesis of an underlying GUT. 
This is the reason why $K \to \pi \nu\bar\nu$ decays, 
which are the best probes of $s \to d$ $\Delta F=1$ short-distance amplitudes, 
play a key role in any extension of the SM containing non-minimal sources 
of flavour symmetry breaking.} Ignoring fine-tuned 
scenarios, $M_\nu \gg  10^{12}$~GeV is excluded by the present constraints 
on quark FCNC transitions. Independently from the value of $M_\nu$, 
deviations from the standard MFV pattern can appear both in $K$ and in $B$ physics
for $\tan\beta \gsim m_t/m_b$. 
\item 
Contrary to the non-GUT MFV framework for the lepton sector,  
the rate for $\mu \to e \gamma$ and other LFV decays cannot be 
arbitrarily suppressed by lowering the mass of the heavy  $\nu_R$. 
This fact can easily be understood by noting that 
the GUT group allows also $M_\nu$-independent contributions 
to LFV decays proportional to the 
quark Yukawa couplings. The latter become competitive 
for $M_\nu \lsim 10^{12}$ GeV and their contribution is such that for 
$\Lambda \lsim 10$ TeV  the  $\mu \to e \gamma$ rate is above 
$10^{-13}$ (i.e.~within the reach of  MEG).
\item 
Within this framework improved experimental searches  
on $\tau \to \mu \gamma$ and $\tau \to e \gamma$ are 
a key tool: they are the best observables to discriminate 
the relative size of the non-GUT MFV contributions with 
respect to the GUT ones. In particular, if the quark-induced terms turn out 
to be dominant, the  $\mathcal{B}(\tau\to\mu\gamma)/\mathcal{B}(\mu\to e\gamma)$
ratio could reach values of $\cO(10^{-4})$, allowing $\tau\to\mu\gamma$ 
to be just below the present exclusion bounds. 
\end{itemize}

\section{The MFV hypothesis in the MSSM}
\label{sec:MSSM}

Low-energy supersymmetry provides an elegant solution to the gauge hierarchy problem, allows for precision gauge coupling unification and accomodates candidates for dark matter (see \cite{Martin:1997ns} for a comprehensive introduction). These virtues make the minimal supersymmetric extension of the SM (MSSM) one of the most extensively studied avenues of new physics and suggest supersymmetric partners to be present around the TeV scale (see \cite{Ellis:2007fu,Buchmueller:2011aa,Buchmueller:2011sw} for recent global fits of SUSY scenarios).
However, it is worth to recall that the adjective {\em minimal}
in the MSSM acronym refers to the particle content of the model 
and not to its flavour structure. In general, the MSSM contains a 
huge number of free parameters, most of them related 
to the flavour structure of the model (sfermion masses
and trilinear couplings). As long as we are ignorant about the mechanism of SUSY breaking, some assumptions about the soft SUSY breaking terms are necessary to avoid excessive flavour violation. The most restrictive -- and, arguably, most successful -- assumption of this type is MFV.

Since the new degrees of freedom (in particular the squark fields) 
have well-defined transformation
properties under the quark-flavour group ${\mathcal G}_q$,
the MFV hypothesis can easily be implemented in the MSSM following the general rules outlined in 
Sect.~\ref{sec:mfv}: 
we need to consider all possible interactions compatible 
with i) softly-broken supersymmetry; ii) the breaking of 
${\mathcal G}_q$ via the spurion fields $Y_{U,D}$. 
This allows to express the squark mass terms and 
the trilinear quark-squark-Higgs couplings 
as follows~\cite{Hall:1990ac,D'Ambrosio:2002ex}:
\bea
{\tilde m}_{Q_L}^2 &=& {\tilde m}^2 \left( a_1 \identity 
+b_1 Y_U Y_U^\dagger +b_2 Y_D Y_D^\dagger \right. \no\\
&& \left. +b_3 Y_D Y_D^\dagger Y_U Y_U^\dagger
+b_4 Y_U Y_U^\dagger Y_D Y_D^\dagger +\ldots
 \right), \qquad 
\label{prima}\\
{\tilde m}_{U_R}^2 &=& {\tilde m}^2 \left( a_2 \identity 
+b_5 Y_U^\dagger Y_U +\ldots \right)~,\\
{\tilde m}_{D_R}^2 &=& {\tilde m}^2 \left( a_3 \identity 
+b_6 Y_D^\dagger Y_D +\ldots \right)~,\\
A_U &=& A\left( a_4 \identity 
+b_7 Y_D Y_D^\dagger +\ldots \right) Y_U~,\\
 A_D &=& A\left( a_5 \identity 
+b_8 Y_U Y_U^\dagger +\ldots \right) Y_D~,
\label{ultima}
\eea
where the dimensional parameters $\tilde m$ and $A$ 
set the overall scale of the soft-breaking terms.
In Eqs.~(\ref{prima})--(\ref{ultima}) we have explicitly shown  
all independent flavour structures which cannot be absorbed into 
a redefinition of the leading terms (up to tiny contributions 
quadratic in the Yukawas of the first two families). 
When $\tan\beta$ is not too large and the bottom Yukawa coupling 
is small, the terms quadratic in $Y_D$ can be dropped.

In a bottom-up approach, the adimensional coefficients 
$a_i$ and $b_i$ in Eqs.~(\ref{prima})--(\ref{ultima})
should be considered as free parameters of the model.
Note that  this structure is 
renor\-ma\-lization-group invariant: the value of 
$a_i$ and $b_i$ change according to the Renormalization Group (RG) 
flow, but the general structure of Eqs.~(\ref{prima})--(\ref{ultima})
is unchanged, as has been demonstrated explicitly in Refs. \cite{Paradisi:2008qh,Colangelo:2008qp}. This is not the case if the $b_i$ are set to zero
(corresponding to the so-called hypothesis of flavour universality).
If this hypothesis is set as initial condition 
at some high-energy scale $M$, then non vanishing 
$b_i \sim (1/4\pi)^2 \ln M^2/ {\tilde m}^2$ are 
generated by the RG evolution. 
This is for instance what happens in  models with gauge-mediated 
supersymmetry breaking~\cite{Dine:1993yw,Dine:1994vc,Giudice:1998bp}, 
where the scale $M$ is identified
with the mass of the hypothetical messenger particles. 

Using the soft terms in Eqs.~(\ref{prima})--(\ref{ultima}), the physical
squark $6\times 6$ mass matrices, after electroweak breaking, 
assume the form shown in Table~\ref{tab:SUSYM}.
\begin{table*}
\beqa
&&   \!\!\!\!\!\! 
{\tilde M}_U^2 = 
\left(
\ba{cc}
{\tilde m}_{Q_L}^2+Y_UY_U^\dagger v_U^2+\left( \frac{1}{2} -\frac{2}{3}\sw
\right) M_Z^2\cos 2\beta & \left( A_U -\mu Y_U \cot \beta \right) v_U \\
 \left( A_U -\mu Y_U \cot \beta \right)^\dagger v_U & \!\!\!\!\!\!
{\tilde m}_{U_R}^2+Y_U^\dagger Y_U v_U^2+\frac{2}{3}\sw
 M_Z^2\cos 2\beta 
\ea
\right)~, \no \\
&&  \!\!\!\!\!\! 
 {\tilde M}_D^2  =
\left(
\ba{cc}
{\tilde m}_{Q_L}^2+Y_DY_D^\dagger v_D^2-\left( \frac{1}{2} -\frac{1}{3}\sw
\right) M_Z^2\cos 2\beta & \left( A_D -\mu Y_D \tan \beta \right) v_D \\
 \left( A_D -\mu Y_D \tan \beta \right)^\dagger v_D & \!\!\!\!\!\!
{\tilde m}_{D_R}^2+Y_D^\dagger Y_D v_D^2-\frac{1}{3}\sw
 M_Z^2\cos 2\beta 
\ea
\right)~. \no
\eeqa
\caption{Squark Mass matrices in MFV. Here $\mu$ is the higgsino mass parameter and 
$v_{U,D}= \langle H_{U,D}\rangle$~ ($\tan\beta =v_U/v_D$).
\label{tab:SUSYM} }
\end{table*}
The eigenvalues of these mass matrices 
are not degenerate; however, the mass splittings are tightly 
constrained by the specific (Yukawa-type) symmetry-breaking pattern. 

If we are interested only in low-energy processes, we can integrate 
out the supersymmetric particles at one loop and project this 
theory onto the general EFT discussed in the previous sections. 
In this case, the coefficients of the dimension-six effective operators 
written in terms of SM and Higgs fields (see Table~\ref{tab:MFV}) 
are computable in terms of the supersymmetric soft-breaking parameters.
We stress that if $\tan\beta \gg 1$ (see Sect.~\ref{sec:largetanb})
and/or if $\mu$ is large enough~\cite{Altmannshofer:2007cs}, 
the relevant operators thus obtained go beyond the restricted 
basis of the CMFV scenario~\cite{Buras:2003jf}. 
We also emphasize that the integration of the supersymmetric degrees 
of freedom may lead to sizable modifications of the renormalizable operators
and, in particular, of the effective Yukawa interaction. 
As a result, in an effective field theory 
with supersymmetric degrees of freedom, the relations between $Y_{U,D}$ 
and the physical quark masses and CKM angles are potentially modified.
As already pointed out in Sect.~\ref{sec:largetanb}, this effect 
is particularly relevant in the large $\tan\beta$ regime.

Comparing the resulting operators to the bounds in Table~\ref{tab:MFV}, using the typical effective suppression scale (assuming an overall coefficient $1/\Lambda^2$)
\beq
\Lambda \sim \, 4 \pi \, \tilde m\,,
\eeq
we conclude that
if MFV holds the present bounds on FCNCs do not exclude squarks in 
the few hundred GeV mass range, i.e. in the region currently probed by LHC.
Since the top Yukawa (and at large $\tan\beta$ also the bottom Yukawa) is large, the squark mass matrices in Table~\ref{tab:SUSYM} show that the third generation squark masses, which are currently less constrained by collider searches, can be split from the first two generation ones. A large hierarchy between these mass scales however goes beyond MFV and is the subject of Section~\ref{sec:EffSUSY}.

The flavour signatures of the MFV MSSM are largely identical to the universal MFV predictions discussed in Section~\ref{sec:mfv}. An additional peculiar feature is that the contributions to $\Delta M_{d,s}$ turn out to be always positive \cite{Altmannshofer:2007cs}. In the large $\tan\beta$ regime, the most sensitive channels are $B_s\to\mu^+\mu^-$ and $B\to\tau\nu$, as discussed in Section~\ref{sec:largetanb}.

If CP violating phases beyond the CKM phase are added to the MFV MSSM, new contributions to EDMs and CP asymmetries arise. The irreducible phases in the MFV MSSM (ignoring the lepton sector) are given by the flavour blind phases
\begin{equation}
\text{Arg}(M_i\mu)
\,,~~
\text{Arg}(a_{4,5}\mu)
\,,~~
\text{Arg}(a_{4,5}^*M_i)
\,,
\label{eq:SUSYph}
\end{equation}
where the $M_i$ are gaugino masses and the $a_i$ are defined in Eq.~(\ref{ultima}), as well as the phases in the complex coefficients $b_{3,4}$ and $b_{7,8}$ in Eqs.~(\ref{prima}, \ref{ultima}).

The phases in (\ref{eq:SUSYph}) are strongly constrained by the upper bounds on EDMs and thus have to be tiny in the MFV MSSM. The parameters $b_{3,4}$ are suppressed unless $\tan\beta$ is large. If the coefficients $b_{7,8}$ are complex, one essentially obtains complex third generation trilinear couplings and real ones for the first two generations. This setup has interesting implications for electric dipole moments and CP violation in $\Delta B=1$ decays \cite{Altmannshofer:2008hc}. However, the $b_{7,8}$ and the $a_{4,5}$ mix under renormalization, so this scenario is not RG invariant and typically suffers from excessive EDMs if imposed at a high scale (unless only $b_7$ is real and $\tan\beta$ small) \cite{Paradisi:2009ey}.

Interestingly, even with flavour blind phases, a significant modification of the SM predictions for the $B_s$ or $B_d$ mixing phases is impossible in the MFV MSSM, even at large $\tan\beta$, once $\Delta F=1$ constraints are taken into account \cite{Altmannshofer:2009ne}. This restriction can be avoided only by introducing non-MFV sources of flavour violation or by extending the MSSM, e.g. by introducing higher-dimensional operators respecting MFV and carrying flavour blind phases \cite{Altmannshofer:2011rm,Altmannshofer:2011iv}.

\section{Flavour symmetry breaking with split families}
\label{sec:EffSUSY}

The MSSM with very heavy superpartners of the first two generation quarks, is a long-standing alternative to MFV. While the solution to the gauge hierarchy problem requires mostly the third generation squarks to be light, the tight constraints from FCNCs are loosened in presence of a squark mass hierarchy~\cite{Dimopoulos:1995mi,Cohen:1996vb}.
Moreover, such spectrum is favoured by LHC sparticle searches: while the bounds on first generation squark masses are approaching a TeV, the third generation ones can still be significantly lighter \cite{ATLAS-CONF-2011-098}.
However, a hierarchical spectrum is not enough to suppress flavour violation for a generic soft SUSY breaking sector \cite{Giudice:2008uk}. Combining split-family SUSY with MFV is thus natural, but not trivial since some of the $U(3)$ factors in $\cG_q$ are broken explicitly by the hierarchical squark mass terms. In the following, we discuss two possible generalizations.

An additional virtue of the split-family framework is the solution to the SUSY CP problem: as mentioned in the introduction, the non-observation of EDMs is a challenge for any SUSY theory and is not addressed by MFV.
Since the observable EDMs are the ones of first generation fermions, the one-loop contributions to these EDMs are strongly suppressed in the case of hierarchical sfermions, where the corresponding superpartners are heavy.

\subsection{\boldmath $U(1)^3_U\times U(3)_{D_R}$}

Motivated by the special role of the top quark Yukawa coupling, in Ref. \cite{Barbieri:2010ar}
a setup was considered based on the following assumptions,
\begin{itemize}
\item among the squarks, only those that interact with the Higgs system via the top Yukawa are light,
\item with only the up-type Yukawas turned on, there is no flavour transition between generations.
\end{itemize}
This corresponds to a flavour symmetry\footnote{The combination of this flavour symmetry with the split-family hypothesis 
 was dubbed ``effective MFV'' in Ref.s \cite{Barbieri:2010ar,Barbieri:2011vn} but we stress that it goes beyond MFV in the sense of section~\ref{sec:mfv}.}
\begin{equation}
U(1)_{U_1} \times U(1)_{U_2} \times U(1)_{U_3} \times U(3)_{D_R}
\end{equation}
broken by the spurion $Y_D$.

In this setup, the first and second generation sfermions are not necessarily degenerate, leading to sizable contributions to $\epsilon_K$. These contributions are ameliorated by the fact that the right-handed sbottom squark is heavy and can be in agreement with the experimental bounds for reasonable values of the parameters. The mixing amplitudes in the $B_d$ and $B_s$ system, on the other hand, remain virtually unaffected.

In Ref.~\cite{Barbieri:2011vn}, it has been demonstrated explicitly that with effective MFV, the EDM problem is solved even for sizable flavour-blind phases. The one-loop contributions to the electron and neutron EDMs are strongly mass-suppressed and under control for first and second generation sfermion masses around 10~TeV, for $O(1)$ phases and moderate $\tan\beta$.
However, two-loop effects become relevant, in particular from Barr-Zee type diagrams not suppressed by heavy sfermion masses. These effects lead to EDM contributions in the ballpark of the present experimental bounds.

\begin{figure*}[t]
\centering
\includegraphics[width=0.32\textwidth]{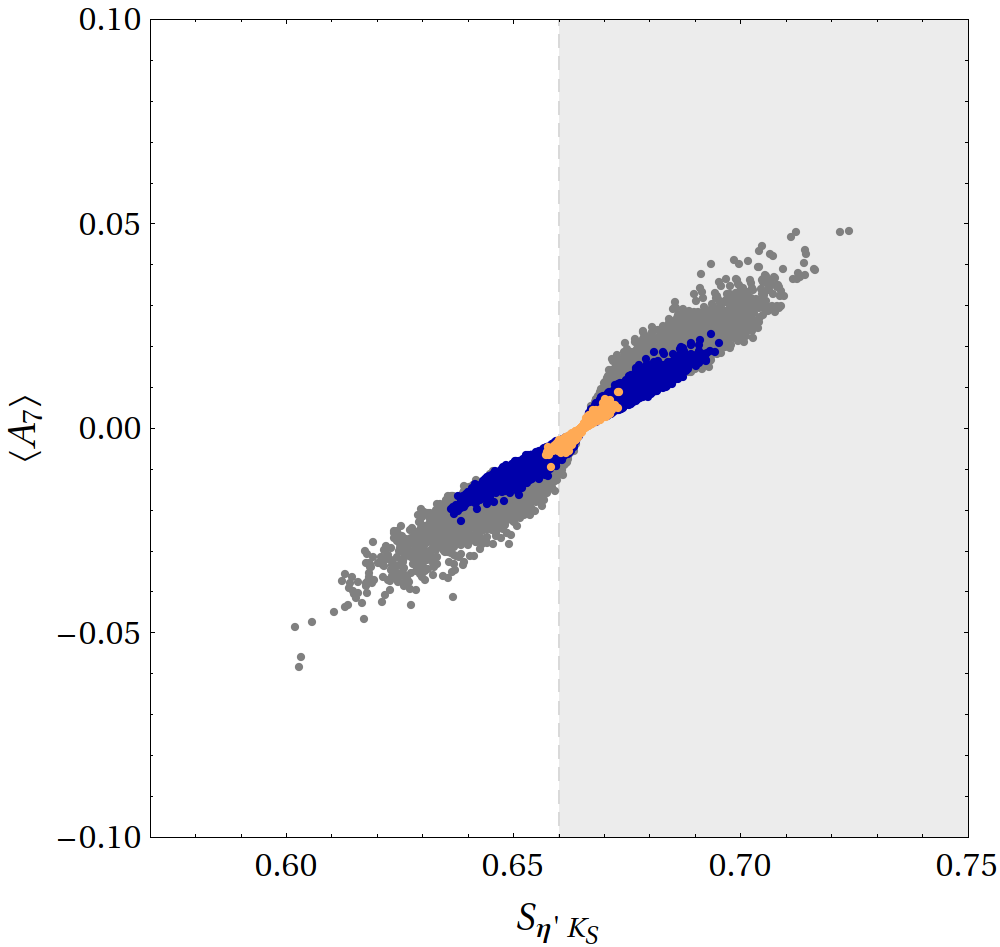}
\includegraphics[width=0.32\textwidth]{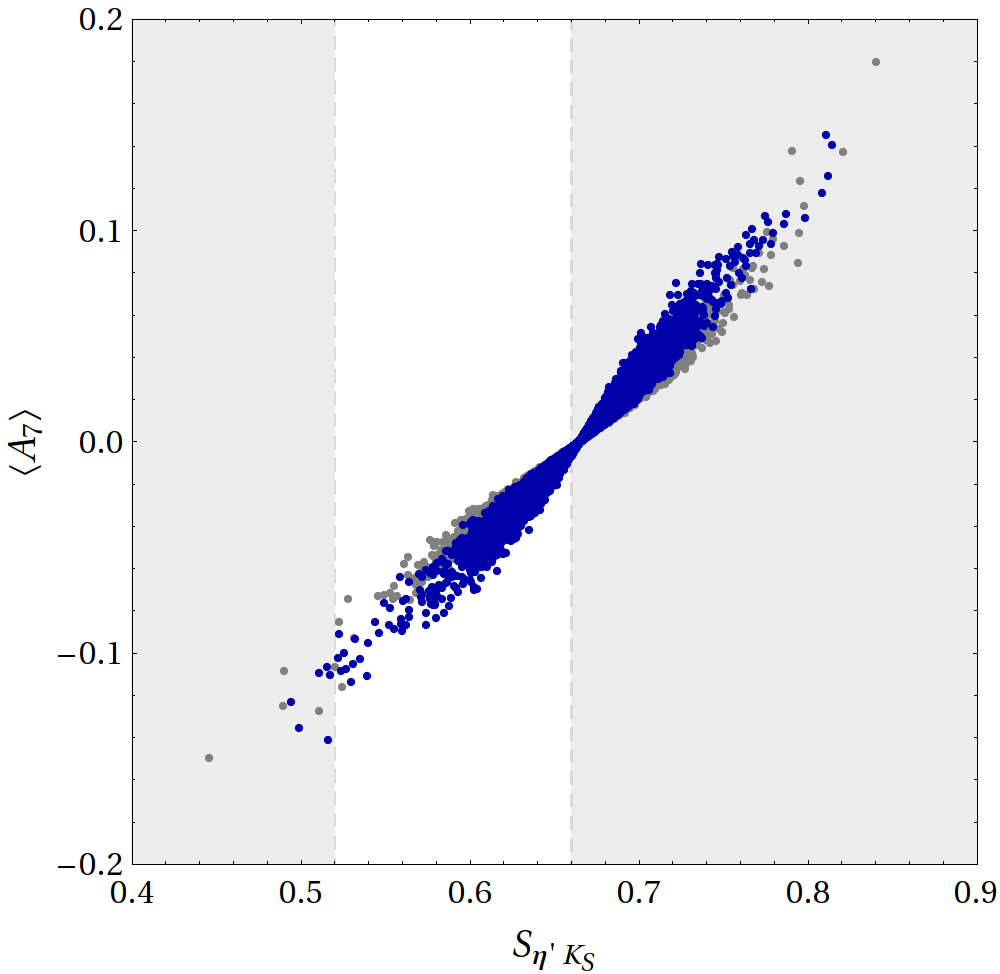}
\includegraphics[width=0.32\textwidth]{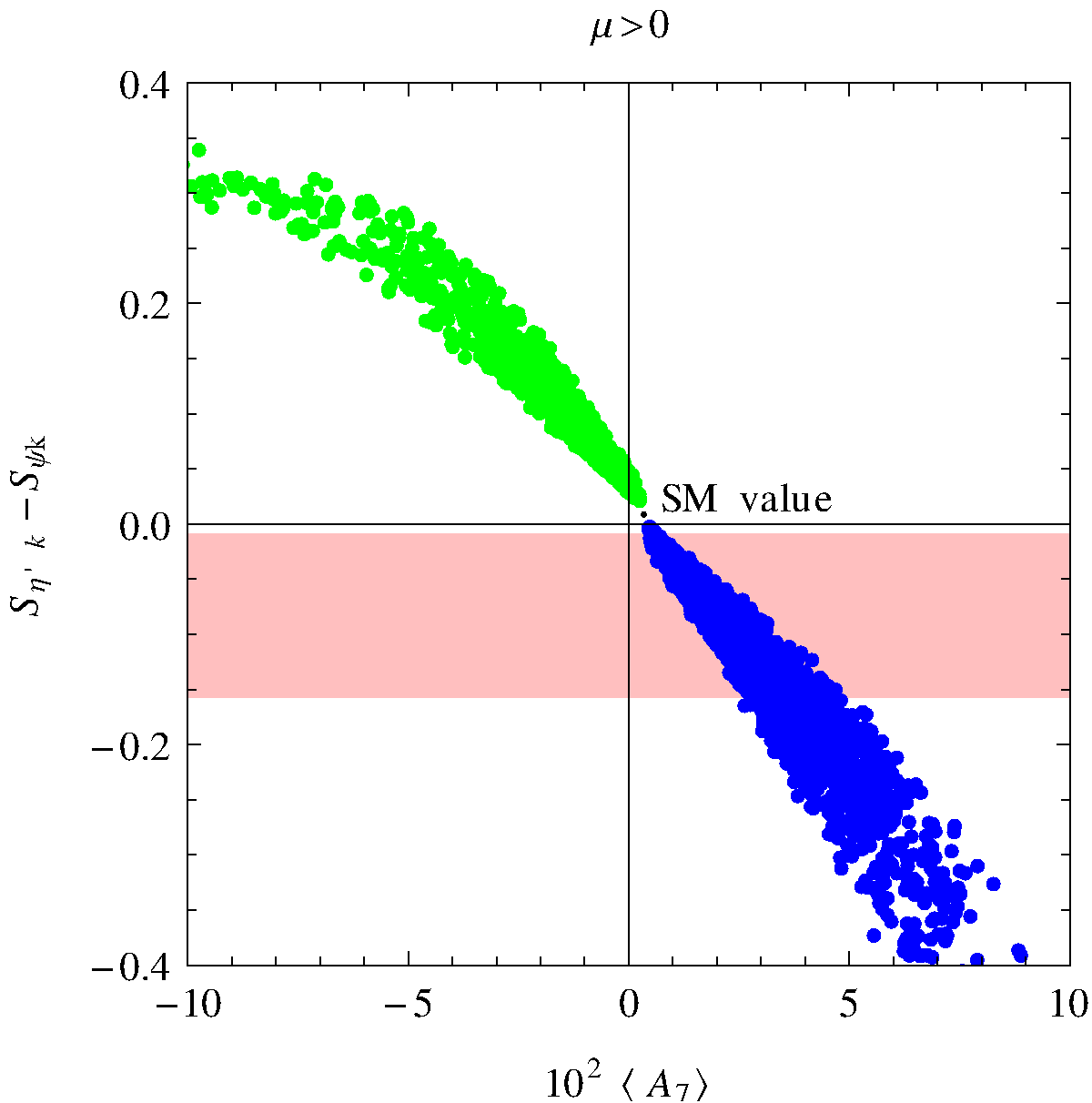}
\caption{%
{\bf Left:} Mixing induced CP asymmetry in $B\to\eta' K_S$ vs. CP asymmetry $\langle A_7 \rangle$ in $B\to K^*\mu^+\mu^-$ for $U(1)^3_U\times U(3)_{D_R}$ with $\tan\beta<5$ and complex $\mu$ term. The gray points are ruled out at 90\% C.L. by the EDM bounds. the orange points have arg$(\mu)<0.2$. The shaded region is disfavoured at $1\sigma$ by $B$ factories.
{\bf Center:} Same as left, but with a real $\mu$ term and complex stop trilinear coupling.
{\bf Right:}
$S_{\eta' K_S}-S_{\psi K_S}$ vs. $\langle A_7 \rangle$ in $U(2)^3$ {\em without} flavour blind phases, with $\tan\beta<10$ and $\mu>0$. The green points have $\gamma<0$, the blue points $\gamma>0$.
The region outside the red band is disfavoured at $1\sigma$ by $B$ factories.
Plots taken from \cite{Barbieri:2011vn,Barbieri:2011fc}.}
\label{fig:df1}
\end{figure*}

Interestingly, even without excessive EDMs, flavour-blind CP violating phases can lead to observable effects in CP asymmetries in $B$ decays. This is because $B$ physics involves the third generation quarks, whose superpartners can be light. As an example, the mixing induced CP asymmetry in $B\to\eta' K_S$ and the CP asymmetry $\langle A_7 \rangle$ in the low dilepton invariant mass region of $B\to K^*\mu^+\mu^-$ are shown in the left and center panels of Fig.~\ref{fig:df1} for a scan with $\tan\beta<5$ and an $O(1)$ flavour blind phase in the $\mu$ term (left) or the stop trilinear term (center).

\subsection{\boldmath $U(2)^3$}
\label{sec:u23}

If all the third generation squarks are heavy, the flavour symmetry in the absence of Yukawa couplings is 
\begin{equation}
\cG_q' = U(2)_{Q_L} \times U(2)_{U_R} \times U(2)_{D_R} 
\,.
\end{equation}
Interestingly, this is also the symmetry preserved by the SM Lagrangian if one neglects the small first and second generation quark masses and the small CKM mixings.
A $U(2)$ symmetry relating the first two generation quark and squark fields has received a lot of attention because it can explain, at least in part, the hierarchies manifest in the Yukawa couplings and at the same time ameliorate the flavour and 
CP problems \cite{Dine:1993np,Pomarol:1995xc,Barbieri:1995uv}. However, a single $U(2)$ acting on left- and right-handed fields turns out not to provide enough protection from flavour violation.

Motivated by these observations, the symmetry $\cG_q'$ has been considered in Ref.~\cite{Barbieri:2011ci} together with an appropriate breaking pattern as an alternative to MFV. 
Analogously to the MFV case, a pair of bi-doublet spurions is introduced, transforming as $\Delta Y_u= (2, \bar{2}, 1)$ and $\Delta Y_d= (2, 1, \bar{2})$.
To allow for communication between the third generation and the first two, at least one additional spurion is needed. The minimal choice compatible with the observed quark masses and mixings is a doublet transforming as $V = (2,1,1)$.

Combining the symmetry breaking terms,
the $3\times 3$ Yukawa matrices in generation space assume the following form,
\begin{align}
Y_u= y_t \left(\begin{array}{cc}
 \Delta Y_u & x_t\,V \\
 0 & 1
\end{array}\right), & &
Y_d= y_b \left(\begin{array}{cc}
 \Delta Y_d & x_b\,V \\
 0 & 1
\end{array}\right),
\label{yukawa}
\end{align}
where $x_t, x_b$ are complex parameters of $\ord{1}$ and the $2\times 2$ matrices $\Delta Y_u$ and $\Delta Y_d$ and the vector $V$ are the small symmetry breaking parameters of $\cG_q'$ with entries of order $\lambda^2$ or smaller, with $\lambda$ the sine of the Cabibbo angle.
Analogous expressions hold for the squark soft mass matrices.

\begin{figure*}[t]
\centering
\includegraphics[width=0.72\textwidth]{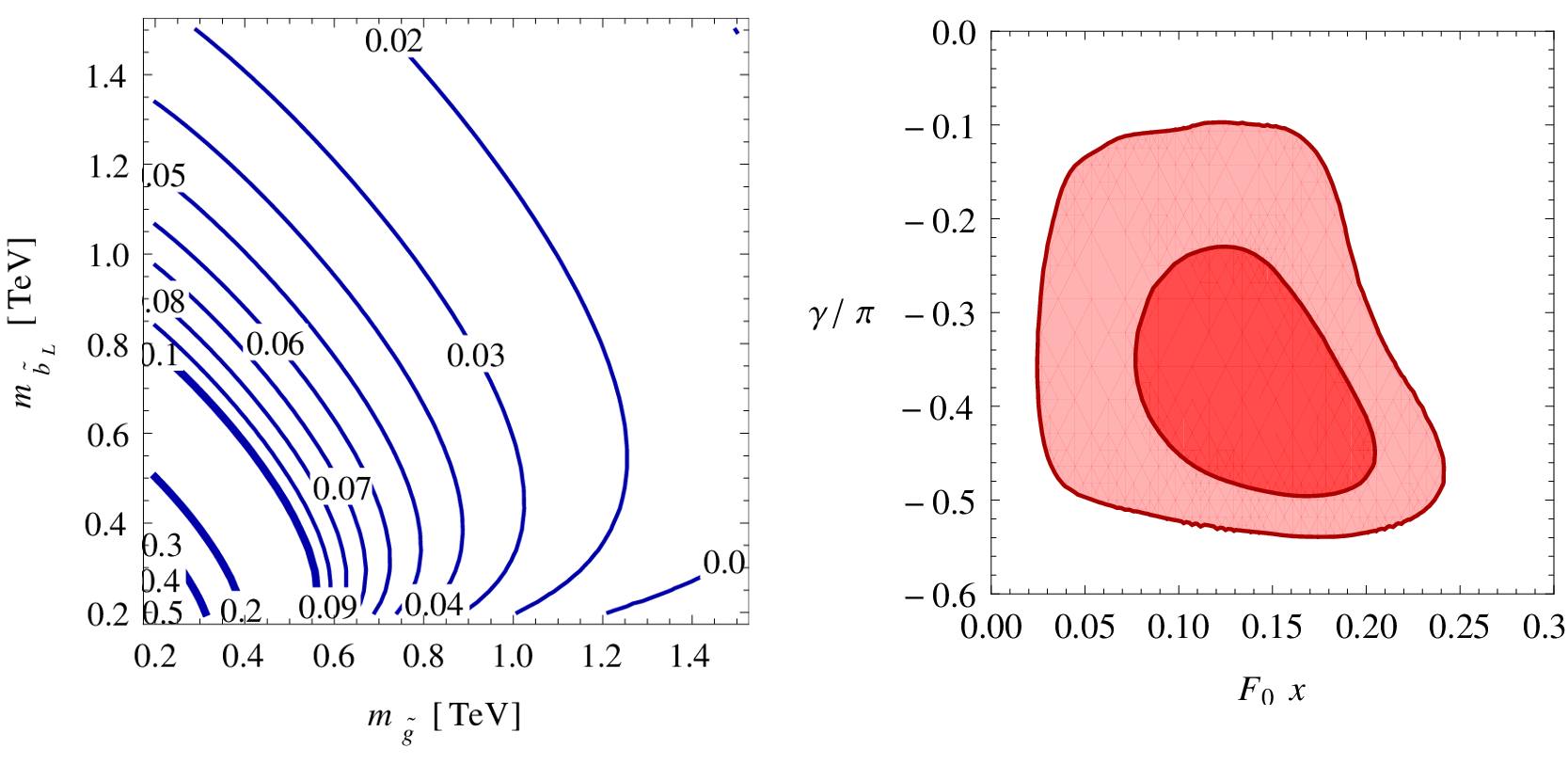}
\caption{%
{\bf Left:}
Values of the gluino-sbottom loop function, $F_0$, controlling  the leading corrections to $\Delta F=2$ observables in 
the MSSM with $U(2)^3$ as a function of gluino and sbottom masses. {\bf Right:} Preferred values of $F_0 \times x$ and the new 
phase $\gamma$, see Eqs.~(\ref{WL})--(\ref{eq:M12s}), as obtained by a fit to present data. Plots taken from \cite{Barbieri:2011ci}. 
}
\label{fig:F0}
\end{figure*}

\begin{figure*}[t]
\centering
\includegraphics[width=\textwidth]{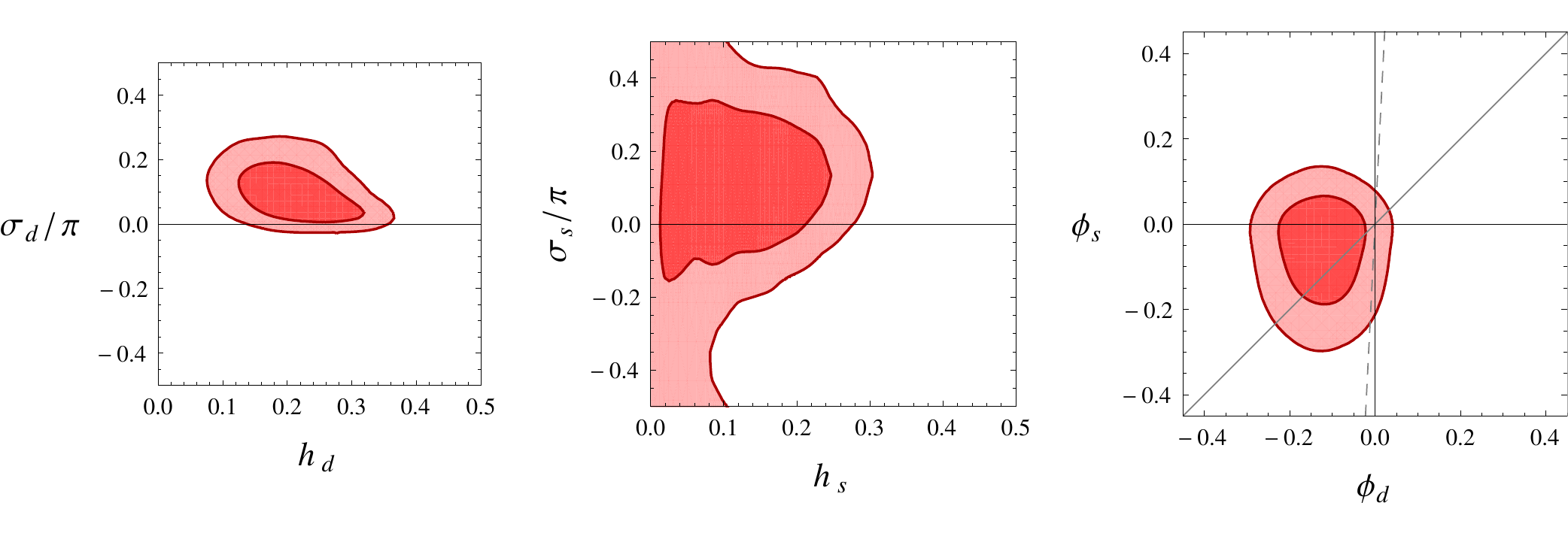}
\caption{Result for a fit to $\Delta F=2$ observables parametrizing the $B_{d,s}$ mixing amplitudes as in Eqs.~(\ref{eq:hsd})--(\ref{eq:hss}).}
\label{fig:hsigma}
\end{figure*}

Of particular phenomenological relevance is the mixing matrix $W_L^d$, diagonalizing the mass matrix of left-handed down-type squarks and appearing in the interaction vertex of left-handed down-type squarks with gauginos. It can be written as
\begin{equation}
W^d_L = \left(\begin{array}{ccc}
 c_d &  s_d  e^{-i(\delta +\phi)}  & -s_d s_L e^{i\gamma} e^{-i(\delta +\phi)}  \\
-s_d e^{i(\delta +\phi)}  &  c_d & -c_d s_L e^{i\gamma}   \\
  0  &  s_L e^{-i\gamma} & 1 \\
\end{array}\right),
\label{WL}
\end{equation}
where $c_d$, $s_d$, $\delta$ and $\phi$ are fixed by the requirement of reproducing the correct quark masses and CKM matrix and $s_L$ and $\gamma$ are a new mixing angle and phase entering $\Delta F=1,2$ transitions and are responsible for the non-MFV effects of $U(2)^3$ in flavour physics.

For example, the absorptive parts of the meson mixing amplitudes in the $K$, $B_d$ and $B_s$ systems can be written as
\begin{align}
M_{12}^K &= (M_{12}^K)_\text{SM}^\text{tt} \left[1+x^2\, F_0\right]+(M_{12}^K)_\text{SM}^\text{tc+cc}
,\label{eq:M12K}\\
M_{12}^{B_d} &= (M_{12}^{B_d})_\text{SM} \left[1+x e^{2i\gamma}\, F_0\right]
,\label{eq:M12d}\\
M_{12}^{B_s} &= (M_{12}^{B_s})_\text{SM} \left[1+x e^{2i\gamma}\, F_0\right]
,\label{eq:M12s}
\end{align}
where $x={s_L^2 c^2_d/|V_{ts}|^2}$ and $F_0$ is a loop function whose numerical value, assuming the dominance of gluino contributions, value is shown in the left panel of Fig.~\ref{fig:F0} as a function of the gluino and sbottom masses. We observe that
\begin{itemize}
\item[i.] in all cases the size of the correction is proportional to the 
CKM combination of the corresponding SM amplitude (MFV structure);
\item[ii.] the proportionality coefficient is the same in $B_d$ and 
$B_s$ systems, while it may be different in the $K$ system;
\item[iii.] new CP-violating phases can only appear in the $B_d$ and 
$B_s$ systems (in a universal way);
\item[iv.] the contribution in the $K$ system interferes constructively with the SM.
\end{itemize}
i.--iii. are model-independent consequences of the $U(2)^3$ symmetry; ii.~and iii.~have also been discussed in  the literature in the context of MFV in the large $\tan\beta$ limit and GMFV~\cite{D'Ambrosio:2002ex,Kagan:2009bn}, which can be viewed as a special cases of a $U(2)^3$ symmetry.
A universal $B_d$ and $B_s$ mixing phase is also predicted in a two Higgs doublet model with MFV \cite{Buras:2010mh,Buras:2010zm} and some MFV extensions of the MSSM \cite{Altmannshofer:2011rm,Altmannshofer:2011iv} with flavour blind phases in the Higgs potential.
iv. is a prediction of $U(2)^3$ in supersymmetry with dominance of gluino contributions to the amplitudes. Interestingly, iv. leads to an unambiguously positive contribution to $\epsilon_K$, as is preferred by the data. Also the non-standard contribution to the $B_d$ mixing phase is welcome in view of tensions in the SM CKM fit among $\epsilon_K$, $\sin2\beta$ and $\Delta M_d/\Delta M_s$.

The right panel of Fig.~\ref{fig:F0} shows the result of a global fit of the CKM matrix, using the 4 Wolfenstein parameters as well as $x$, $F_0$ and $\gamma$ as inputs to Eqs. (\ref{eq:M12K})--(\ref{eq:M12s}) and all relevant observables from tree-level and loop-induced processes as constraints (for details see \cite{Barbieri:2011ci}). One can observe a non-zero value favoured for the NP phase $\gamma$, driven by the tensions in the SM CKM fit. $F_0$ and $x$ are not well constrained separately, but sub-TeV masses for the sbottom and gluino are favoured by the fit.

\definecolor{one}{rgb}{0.7,0.7,0.7} 
\definecolor{two}{rgb}{0.75,0.35,0.35} 
\definecolor{three}{rgb}{0.8,0,0} 
\newcommand{\three}{{\color{three}\Large$\star\!\star\!\star$}}
\newcommand{\two}{{\color{two}\large$\star\star$}}
\newcommand{\one}{{\color{one}$\star$}}
%
\begin{table*}[t]
\addtolength{\arraycolsep}{4pt}
\renewcommand{\arraystretch}{1.4}
\centering
\begin{tabular}{|l|ccc|ccc|}
\hline
& $U(3)^3$ & $U(1)^3\!\times\!U(3)$ & $U(2)^3$
& $U(3)^3$ & $U(1)^3\!\times\!U(3)$ & $U(2)^3$
\\
& \multicolumn{3}{c|}{no flavour-blind phases} & \multicolumn{3}{c|}{with flavour-blind phases} \\
\hline
$\epsilon_K$& \two & \three & \two & \two & \three & \two
\\
$S_{\psi\phi}$ & \one & \one & \three & \one & \one & \three
\\
$S_{\phi K_S}$ & \one & \one & \three & \three & \three & \three
\\
$B\to K^*\ell^+\ell^-_{\rm \ [CPV\  asym.]}$ & \one & \one & \two & \two & \two & \two
\\
EDMs & \one & \one & \one & \three & \three & \three
\\
\hline
\end{tabular}
\renewcommand{\arraystretch}{1}
\caption{%
Maximal possible size of new physics effects for different flavour symmetries (and the associated flavour symmetry breaking as discussed in the text): only small effects possible (1 star), moderate effects possible (2 stars) or large effects possible (3 stars).}
\label{tab:DNA}
\end{table*}

Recently, the LHCb collaboration has measured the $B_s$ mixing phase in $B_s\to J/\psi\phi$ and $B_s\to J/\psi f_0$ decays and found agreement, within errors, with the SM prediction \cite{LHCb-CONF-2011-056}. To confront the $U(2)^3$ prediction of a universal modification of $B_d$ and $B_s$ mixing phases with the data, we performed a further fit of the CKM matrix, allowing for a general modification of the $\Delta B=2$ amplitudes,
\begin{align}
M_{12}^{B_d} &= (M_{12}^{B_d})_\text{SM} \left[1-h_d e^{2i\sigma_d}\right]
\label{eq:hsd}
,\\
M_{12}^{B_s} &= (M_{12}^{B_s})_\text{SM} \left[1-h_s e^{2i\sigma_s}\right]
,
\label{eq:hss}
\end{align}
assuming for simplicity no NP in $K$ mixing.
As constraints, we used  the same observables as above, but in addition the data on the $B_s$ mixing phase from LHCb, CDF \cite{CDF-NOTE-10206} and D0 \cite{D0Note-6098-CONF}. The fit results are shown in Fig.~\ref{fig:hsigma}. The left and center plot show a preference for a non-standard contribution to $B_d$ mixing, while $B_s$ mixing is compatible with the SM at 68\% C.L.. The right plot shows the preferred region (cf. also \cite{Altmannshofer:2011iv}) for the NP phases defined as 
\begin{equation}
\phi_q = \text{arg}\left(1-h_q e^{2i\sigma_q}\right).
\end{equation}
In this plot, the origin is the SM point, while the $U(2)^3$ prediction $\phi_d=\phi_s$ is shown as a solid line.
As already discussed, the $\phi_d=\phi_s$  prediction holds beyond the hypothesis of supersymmetry with split families. It was found
for instance in two Higgs doublet models with MFV \cite{Buras:2010mh,Buras:2010zm} and 
in the MFV MSSM with an extended Higgs sector with flavour blind phases in the Higgs potential~\cite{Altmannshofer:2011iv}.
However, in the latter models a large breaking of the Peccei-Quinn symmetry and 
flavour-blind phases only in the effective Yukawa interaction lead instead to the prediction $\phi_d=(m_d/m_s)\phi_s$ \cite{Altmannshofer:2011iv,Buras:2010mh,Buras:2010zm}, which is shown as a dashed line. One can see that current data favour the $U(2)^3$ prediction over the latter prediction as well as the SM.

Contrary to the $\Delta F = 2$ sector, where the pattern of deviations from the SM is unambiguously
dictated by the $U(2)^3$ symmetry, the predictions of $\Delta F = 1$ observables are more model
dependent. Within supersymmetry with a $U(2)^3$ symmetry, it has been shown in \cite{Barbieri:2011fc} that visible effects can be generated in particular in the mixing induced CP asymmetries in $B\to\phi K_S$ and $B\to\eta' K_S$, in the CP asymmetry $\langle A_7 \rangle$ in $B\to K^*\mu^+\mu^-$ and in the direct CP asymmetry in $B\to X_s\gamma$. While these processes are also the golden modes in MFV or ``effective MFV'' with flavour blind phases, they can be generated in $U(2)^3$ even in the absence of flavour blind phases by means of the new phase~$\gamma$. In the right panel of Fig.~\ref{fig:df1}, we show the difference $S_{\eta' K_S}-S_{\phi K_S}$ vs. $\langle A_7 \rangle$ for a scan with $\tan\beta<10$ and positive $\mu$.
While the contribution to the mixing phase in (\ref{eq:M12d}) affects $S_{\eta' K_S}$ and $S_{\phi K_S}$ in the same way, the $\Delta F=1$ penguin contributions modify only $S_{\eta' K_S}$.

\section{Conclusions}

As anticipated in the introduction, the MFV hypothesis and its variations discussed here do not represent a complete 
answer to the flavour problem. They  provide only an efficient answer to the question of why we have not seen 
large deviations from the SM in flavour-changing processes so far, under the assumption of new physics
close to the TeV scale. However, they also provide an efficient tool to derive phenomenological 
predictions about deviations from the SM in  flavour-changing processes that, when confronted with data,
could lead to model-independent conclusions about the irreducible sources of flavour symmetry breaking 
accessible at low energies. 

In table~\ref{tab:DNA} we summarise the possible maximal deviations from the SM, taking into account theoretical errors and 
near-future experimental sensitivity, in a series of particularly clean CP-violating observables. We compare 
the MFV expectations to those of the two other flavour symmetries discussed in section~\ref{sec:EffSUSY}, distinguishing the case where 
CP violation originates from the Yukawa couplings only and the case with additional flavour-blind phases
(for more quantitative details see \cite{Isidori:2010kg,Altmannshofer:2008hc,Barbieri:2011vn,Barbieri:2011ci,Barbieri:2011fc}).
The different pattern of effects can become useful both to distinguish between the
different frameworks, if a significant deviation from the SM expectations is observed in one (or several) of the observables,
or even to proof the existence of  flavour symmetry-breaking terms not related to the Yukawa sector. 

\section*{Acknowledgments}

We thank R.~Barbieri, A.J.~Buras, G.~Giudice, G.~Pe\-rez, and W.~Altmannshofer for useful comments and discussions.
GI acknowledges the support  of the TU~M\"unchen -- Institute for Advanced
Study, funded by the German Excellence Initiative, and the  EU ERC Advanced
Grant FLA\-VOUR (267104).
DMS is supported by the EU ITN ``Unification in the LHC Era'', 
contract PITN-GA-2009-237920 (UNILHC).

\bibliographystyle{My}
\bibliography{mfv}

\end{document}